\documentclass{article}
\usepackage{geometry}
\usepackage{setspace}
\usepackage{titlesec}
\usepackage{graphicx}
\usepackage{booktabs}
\usepackage{enumitem}
\usepackage{xcolor}

\usepackage{hyperref}
\hypersetup{
    colorlinks=true,
    linkcolor=black, % Color for internal links
    citecolor=green, % Color for citations
    urlcolor=blue    % Color for external links
}

\usepackage{authblk}
\usepackage{booktabs}

\usepackage{soul}
\sethlcolor{yellow}

\usepackage{multirow}
\usepackage{color, colortbl}
\definecolor{Gray}{gray}{0.9}
\definecolor{LightCyan}{rgb}{0.88,1,1}
\usepackage[first=0,last=9]{lcg}

\definecolor{LightBlue}{rgb}{0.90,0.95,1}
\definecolor{LightYellow}{rgb}{1, 1, 0.7}
\definecolor{LightGreen}{rgb}{0.7, 0.9, 0.7}

\title{\textbf{When is a Foundation Model a Foundation Model}}
\author[1]{Saghir Alfasly}
\author[1]{Peyman Nejat}
\author[1]{Sobhan Hemati}
\author[1]{Jibran Khan}
\author[1]{Isaiah Lahr}
\author[1]{Areej Alsaafin}
\author[1]{Abubakr Shafique}
\author[2,3]{Nneka Comfere}
\author[2]{Dennis Murphree}
\author[3]{Chady Meroueh}
\author[3]{Saba Yasir}
\author[4]{Aaron Mangold}
\author[5]{Lisa Boardman}
\author[5]{Vijay Shah}
\author[3]{Joaquin J. Garcia}
\author[1]{H.R. Tizhoosh}

\affil[1]{Department of Artificial Intelligence and Informatics, Mayo Clinic, Rochester, MN, USA}
\affil[2]{Department of Dermatology, Mayo Clinic, Rochester, MN, USA}
\affil[3]{Department of Laboratory Medicine and Pathology, Mayo Clinic, Rochester, MN, USA}
\affil[4]{Department of Dermatology, Mayo Clinic, Phoenix, AZ, USA}
\affil[5]{Comprehensive Cancer Center, Mayo Clinic, Rochester, MN, USA}

\date{} % Remove date

\nocite{*}

\begin{document}
\maketitle

\textbf{Abstract --} Recently, several studies have reported on the fine-tuning of foundation models for image-text modeling in the field of medicine, utilizing images from online data sources such as Twitter and PubMed. Foundation models are large, deep artificial neural networks capable of learning the context of a specific domain through training on exceptionally extensive datasets. Through validation, we have observed that the representations generated by such models exhibit inferior performance in retrieval tasks within digital pathology when compared to those generated by significantly smaller, conventional deep networks.

\vspace{0.2in}

% \section{Introduction}
The term \emph{foundation model} (FM) refers to large (i.e., deep) artificial neural networks that, after extensive pre-training and fine-tuning with a very large amount of data, can serve as the backbone (foundation) for a wide range of applications \cite{bommasani2021opportunities, wei2022emergent}. Training FMs to acquire comprehensive and expressive representations of complex data, such as natural language and digital images, requires massive amounts of data. FMs can be further fine-tuned to better understand domain-specific contexts. Understandably, we need an extremely large and diverse corpus of data to train FMs. This commonly includes, for general-purpose FMs, books, articles, websites, and social media posts, among other sources. However, it is important to note that some of these sources may be inaccessible (e.g., due to copyright issues for medical books) or may not be founded on solid medical evidence (e.g., individual opinions and social media) when seeking reliable medical data.

One of the most popular architectures for multimodal foundation models is CLIP (Contrastive Language-Image Pre-training) \cite{radford2021learning}. Developed by OpenAI, CLIP is designed to learn ``joint representations'' of images and their corresponding textual descriptions, enabling tasks like automatic image captioning, image retrieval, and even image generation based on textual descriptions. Leveraging a large dataset containing images and their associated textual descriptions, CLIP learns to closely map similar images and their descriptions in the feature space while effectively discriminating between dissimilar pairs. CLIP has demonstrated impressive capabilities across various tasks, including image classification, object detection, and generating images from textual descriptions \cite{radford2021learning}. Its ability to generalize across diverse datasets and tasks has positioned it as a versatile and powerful tool in the field of AI research and applications.

Recently, Huang et al. published a paper introducing PLIP, which is a fine-tuned version of CLIP using histology images from the pathology communities on Twitter \cite{huang2023visual}. They also conducted comparisons with search techniques for image retrieval, a task of significant interest in the field of digital pathology. Before this, BiomedCLIP had been introduced, once again employing CLIP but fine-tuned on a dataset comprising 15 million online biomedical image-text pairs.

\vspace{0.1in}
Can these foundation models aid in the field of medicine?
\vspace{0.1in}

The adoption of digital pathology, which entails the digitization of tissue slides, has been steadily increasing in recent years \cite{pantanowitz2011review, hanna2022integrating}. Histopathology primarily involves the visual examination of tissue samples using light microscopy. The process of digitizing glass tissue slides is accomplished through specialized scanners that capture high-resolution whole-slide images (WSIs) from tissue samples. Consequently, pathologists can perform visual inspections of tissue morphology on a computer screen, replacing traditional microscope-based examinations. The availability of tissue samples in digital format opens up opportunities for the application of computer vision and artificial intelligence in the field of pathology \cite{heinz2022future}.

WSI-to-WSI comparison stands as one of the pivotal tasks in computational and diagnostic pathology. It has the potential to enable 'patient matching,' paving the way for real-time, evidence-based, and individualized medicine, especially when combined with other data modalities. Although image search technologies, or more precisely, content-based image retrieval, have been available for nearly three decades, WSI-to-WSI matching has only recently become feasible \cite{kalra2020yottixel, kalra2020pan}.
%The main barriers to achieving this capability have included:
%\begin{itemize}
% \item Lack of digitization technology (now addressed by digital scanners).
% \item Lack of anatomically and semantically meaningful tissue representation (now made possible through deep learning) \cite{zhang2023challenges}.
% \item Lack of intelligent 'Divide \& Conquer' algorithms for processing gigapixel WSIs (now addressed by patching algorithms like the 'mosaic' concept \cite{kalra2020yottixel}).
% \end{itemize}

The challenge of accurately matching one patient to another at the WSI level persists and relies heavily on representations or embeddings (i.e., features or attributes) extracted by deep networks from patches of a WSI. Processing the entire WSI in one go is infeasible due to its large dimensions.

\vspace{0.1in}

\textbf{Note}: WSI matching is merely an organized way of combining multiple patch comparisons. Hence, when we perform WSI matching we are still depending on the quality of embeddings of single patches. 

\vspace{0.1in}

Huang et al. did not conduct tests for WSI matching. The search method they tested is a derivative of Yottixel \cite{kalra2020yottixel}, which, due to the ranking concept it employs, cannot perform WSI-to-WSI matching. Furthermore, and more importantly, Huang et al. did not compare PLIP against other deep networks that are not foundational but have been trained specifically for histology/histopathology. Notably, they did not include KimiaNet \cite{riasatian2021fine} in their comparisons, despite it being trained on all diagnostic slides from the TCGA repository. This raises an urgent question regarding the use of foundation models: \textbf{\textit{Can FMs trained on datasets other than high-quality clinical data provide the best-of-breed embeddings (i.e., features) necessary to support downstream tasks like patient matching?}}

\begin{figure}[htbp]
    \centering
    \includegraphics[width=0.95\textwidth]{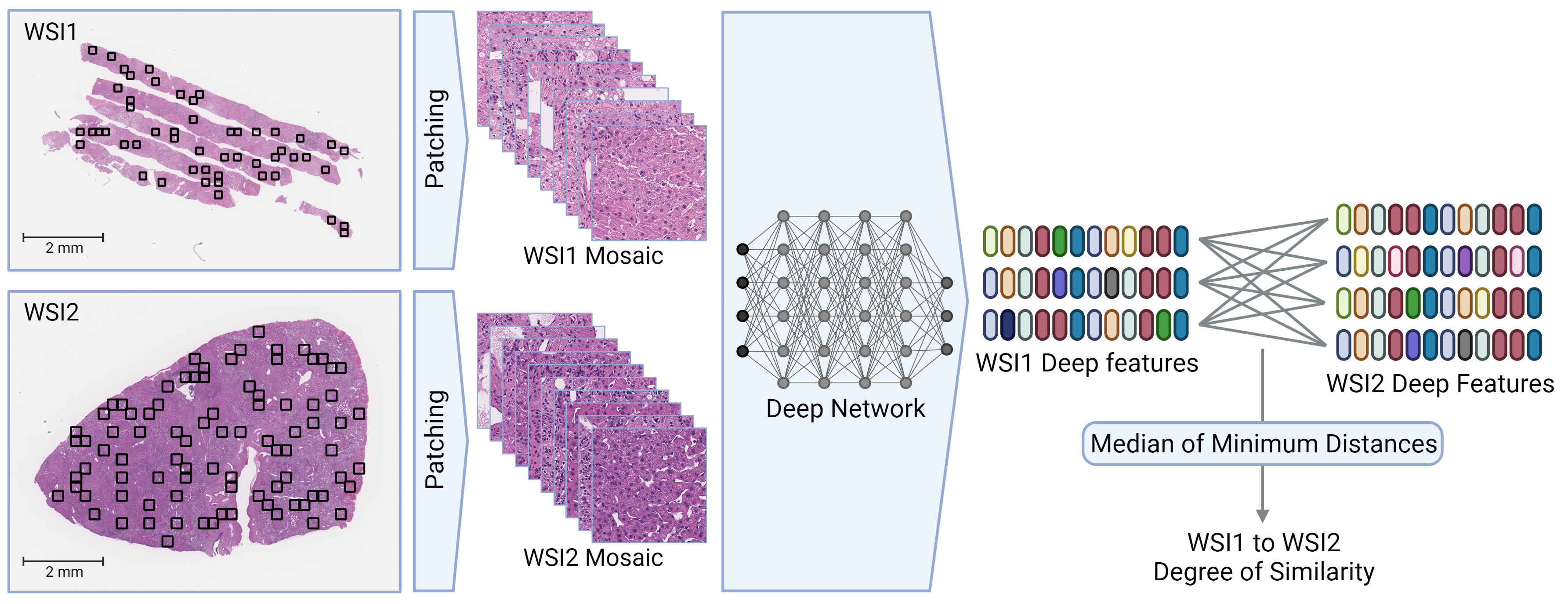} % Replace with your image file name and path
    \caption{\textit{Patching to build a mosaic and median-of-min distance measurements enables WSI-to-WSI comparison (patient matching) \cite{kalra2020yottixel}. Any deep network (conventional or foundational) can be used to extract embeddings (deep features).  [The number of patches, and the number and length of deep feature vectors do not correspond to actual numbers in WSI for the sake of less crowded visualization]}}
    \label{fig:overview}
\end{figure}

% \section{Methods}
To answer this urgent question we examined three FMs that can handle digital images:
\begin{itemize}
    \item CLIP (trained with 400 million image-caption pairs) is a collection of multiple deep models with approximately 100 million parameters \cite{radford2021learning}.
    \item BiomedCLIP (fine-tuned CLIP with 15 million biomedical image-text pairs scraped from online sources) \cite{zhang2023large}.
    \item PLIP (fine-tuned CLIP with 208,414 histopathology patches scraped from Twitter) \cite{huang2023visual}.
\end{itemize}

To provide a comprehensive context, we also employed a simple conventional deep network architecture (i.e., KimiaNet which is based on the DenseNet architecture) and a Transformer-based architecture (DinoSSLPath \cite{Kang_2023_CVPR} which is based on a small vision transformer). Both models have been trained on TCGA repository in which KimiaNet \cite{riasatian2021fine} was trained in a supervised manner, whereas DinoSSLPath was trained in a self-supervised manner. We extracted tissue features at the patch-level using CLIP, BiomedCLIP, PLIP, DinoSSLPath, and KimiaNet. Subsequently, we utilized Yottixel to conduct WSI-to-WSI matching based on the extracted features \cite{kalra2020yottixel, kalra2020pan}. Common convolutional neural networks like KimiaNet are expected to be inferior to foundation models like CLIP and its derivatives when it comes to ``representation'' learning. Foundation models are expected to excel in extracting optimal features for representing input data, thanks to their more complex structure and being trained with substantially more data. Our primary focus was on achieving ``zero-shot retrieval'' using similarity measurements for WSI-to-WSI matching as a downstream task (see Figure \ref{fig:overview}).

\begin{table}[htbp]
    \centering
    \caption{The results of patient matching for classification, subtyping and/or grading for the four internal and two public datasets. We used the ``mosaic'' patching method and the ``median-of-minimum'' matching method introduced by the Yottixel search engine, to find the most similar patients. The validation was done using the ``leave-one-patient-out'' method. The embeddings (i.e., representations or features) were provided by CLIP, BiomedCLIP, PLIP, DinoSSLPath, and KimiaNet. We employed the average of F1 Scores to include both precision and recall values using MV@k (majority vote among top-k retrieved WSIs). The results in green are the best results, whereas the second-best results are highlighted in yellow. \vspace{0.05in}}
    
    \begin{tabular}{lcccc|cc}
    \toprule
    \addlinespace
    \multirow{2}{*}{\textbf{Internal Datasets}}  \\
    \cline{3-7} \addlinespace
    & & PLIP & BiomedCLIP & CLIP & DinoSSLPath & KimiaNet \\
    \midrule
    \midrule
    Breast & Top 1 & 45\% &  39\% & 33\% &\cellcolor{LightYellow}55\%& \cellcolor{LightGreen} 56\% \\\toprule
    \multirow{3}{*}{Liver} & Top 1 &  58\% & 59\% &  52\% &\cellcolor{LightGreen}65\%& \cellcolor{LightYellow} 62\% \\
    & MV@3 & \cellcolor{LightYellow} 68\% &  \cellcolor{LightYellow} 68\% & 50\% &\cellcolor{LightGreen}69\%&  67\% \\
    & MV@5 & 59\% &  64\% & 56\% &\cellcolor{LightGreen}74\%& \cellcolor{LightYellow} 65\% \\\toprule
    \multirow{3}{*}{Skin} & Top 1 &  63\% & 61\% &  62\% &\cellcolor{LightYellow} 71\%& \cellcolor{LightGreen} 78\% \\
    & MV@3 &  65\% & 62\% &  65\% &\cellcolor{LightYellow} 68\% &\cellcolor{LightGreen} 70\% \\
    & MV@5 & \cellcolor{LightYellow} 67\% & \cellcolor{LightYellow} 67\% & \cellcolor{LightYellow} 67\% &66\%& \cellcolor{LightGreen} 69\% \\\toprule
    \multirow{3}{*}{Colorectal} & Top 1 & \cellcolor{LightYellow} 59\% & 54\% &  54\% &\cellcolor{LightGreen}60\%& \cellcolor{LightGreen} 60\% \\
    & MV@3 & \cellcolor{LightYellow} 60\% &  58\% &  55\% &\cellcolor{LightGreen}61\%& \cellcolor{LightGreen} 61\% \\
    & MV@5 & \cellcolor{LightYellow} 61\% & 57\% & 50\% &\cellcolor{LightGreen}63\%&  60\% \\  \rowcolor{LightBlue}\hline
    Total F1 Score & & 60\% $\pm$ 6\% &  59\% $\pm$ 8\% & 55\% $\pm$ 10\% & \textbf{ 65\% $\pm$ 5\%}& \textbf{ 65\% $\pm$ 6\%} \\\hline\hline \addlinespace
    \multirow{2}{*}{\textbf{Public Datasets}}  \\
    \cline{3-7} \addlinespace
    & & PLIP & BiomedCLIP & CLIP & DinoSSLPath & KimiaNet \\
    \midrule
    DigestPath {\tiny \textbf{CTS}} & MV@5 & 84.1\% & 87.2\%&86.9\%     &\cellcolor{LightGreen} 91.5\% & \cellcolor{LightYellow} 89.1\% \\
    DigestPath {\tiny \textbf{SRCC}}& MV@5 &90.8\% & 95.4\%&96.1\%    &\cellcolor{LightGreen} 98.8\% & \cellcolor{LightYellow} 98.0\%\\\toprule
    WSSS4LUAD & MV@5 & 47.4\% & \cellcolor{LightYellow} 53.6\%&48.6\%   &\cellcolor{LightGreen} 56.7\% & 51.3\% \\
    \midrule
    \rowcolor{LightBlue} Total F1 Score & & 79.1$\pm$23 & 78.7$\pm$22 & 77.2$\pm$25 & \textbf{82.3$\pm$22} & \textbf{79.5$\pm$24} \\
    \bottomrule
    \end{tabular}
    \label{tab:results}
\end{table}

We used four internal to examine the quality of representations:

\begin{itemize}
    \item \textbf{Breast Epithelial Tumors} (73 patients) [16 subtypes: 'Adenoid Cystic Carcinoma', 'Adenomyoepthelioma', 'DCIS', 'LCIS', 'Microglandular Adenosis', etc.]
\item \textbf{Fatty Liver Disease} (324 patients) [3 classes: Normal tissue, Non-alcoholic steatohepatitis, alcoholic steatohepatitis]
\item \textbf{Cutaneous Squamous Cell Carcinoma} (660 patients) [4 classes: Normal tissue, well/moderately/poorly differentiated]
\item \textbf{Colorectal Polyps} (209 patients) [3 classes: Cancer Adjacent Polyp, Non-recurrent Polyp, Recurrent Polyp]
\end{itemize}

We also tested the models on two public datasets.  The results of WSI-to-WSI matching are reported in Table \ref{tab:results}. Upon analyzing the results, it becomes evident that both BiomedCLIP and PLIP have enhanced the performance of the original CLIP (when applied to our internal data), which aligns with the common expectation of what fine-tuning any deep model, whether foundational or not, should achieve. The surprising observation is that KimiaNet and DinoSSLPath, relatively standard CNN/transformer models with fewer parameters and training data, provide superior representations compared to CLIP architectures with approximately 100 million parameters. While there is no empirical or theoretical doubt about the capabilities and reliability of the CLIP topology, it suggests that the data used for fine-tuning, in our case, histopathology data, compensates for this difference. This situation highlights the possibility that models labeled as 'foundational' may struggle to match the performance of 'conventional' models when the latter are trained on more robust data sources.

A foundation model earns its title when it manages to surpass the hurdles of generalization and delivers anatomically and semantically reliable, and thus accurate, image representations in histopathology. This underscores the importance of investing time and resources in the creation of high-quality datasets that can truly unlock the potential of foundation models.

\bibliographystyle{unsrt}
\bibliography{ref}

\end{document}